\documentclass[preprint,showpacs,amssymb,preprintnumbers,amsmath,amssymb,prl,aps]{revtex4-1}
\usepackage{natbib}
\usepackage{latexsym} 
\usepackage{graphicx}
\usepackage{dcolumn}
\usepackage{bm}
\usepackage{float}
\usepackage{color}
\usepackage{soul}
\newcommand{\para}{|\kern-.1em|}

\begin{document}

\title{Doping-dependent band structure of LaAlO$_{3}$/SrTiO$_{3}$ interfaces by soft x-ray polarization-controlled resonant angle-resolved photoemission}
\author{C. Cancellieri}
\email{claudia.cancellieri@psi.ch}
\author{M. L. Reinle-Schmitt}
\author{M. Kobayashi}
\author{V. N. Strocov}
\affiliation{Swiss Light Source, Paul Scherrer Institut, CH-5232 Villigen, Switzerland}
\author{D. Fontaine}
\author{Ph. Ghosez}
\affiliation{Physique Th\'eorique des Mat\'eriaux, Universit\'e de Li\`ege, All\'ee du 6 Ao\^ut 17 (B5), 4000 Sart Tilman, Belgium}
\author{A. Filippetti}
\author{P. Delugas}
\author{V. Fiorentini}
\affiliation{CNR-IOM UOS Cagliari, Dipartimento di Fisica, Universit\`a di Cagliari, SP Monserrato-Sestu km.0.700, 09042 Monserrato (CA), Italy}
\author{P. R. Willmott}
\affiliation{Swiss Light Source, Paul Scherrer Institut, CH-5232 Villigen, Switzerland}

\date{\today }

\begin{abstract}
Polarization-controlled synchrotron radiation was used to map the electronic structure of buried conducting interfaces of LaAlO$_3$/SrTiO$_3$ in a resonant angle-resolved photoemission experiment. A strong dependence on the light polarization of the Fermi surface and band dispersions is demonstrated, highlighting the distinct Ti $3d$ orbitals involved in 2D conduction. Samples with different 2D doping levels were prepared and measured by photoemission, revealing different band occupancies and Fermi surface shapes.
A direct comparison between the photoemission measurements and advanced first-principle calculations carried out for different $3d$-band fillings is presented in conjunction with the 2D carrier concentration obtained from transport measurements.

\end{abstract}

\pacs{79.60.Jv, 73.20.-r, 31.15.A-} \maketitle

Complex-oxide interfaces exhibit a broad spectrum of electronic properties and complex phase diagrams and have thus attracted considerable attention. A particularly interesting example is the appearance of 2-dimensional (2D) conductivity at the interface between the band insulators LaAlO$_3$ (LAO) and SrTiO$_3$ (STO) \cite{ohtomo04, reyren07, caviglia08} above a critical LAO thickness of 3 unit cells (u.c.) \cite{thiel06}. As revealed by \textit{ab-initio} calculations \cite{Popovic2008,Delugas}, the mobile electron charge of this 2-dimensional system (2DES) is confined in conduction bands of $3d$ $t_{2g}$ orbital character extending over a few STO layers from the interface and is thus very different from that of doped STO bulk.  
However, the detailed characteristics of these conduction bands crucially depend on the amount of carriers present at the interface. It follows that a comparison between observed and calculated electronic properties of the interface only makes sense if referred to the same carrier concentration. 

The band structure calculated for 0.5 electrons per u.c. (3.5$ \times 10^{14}$ e/cm$^2$), that is the value needed to suppress the ``polar catastrophe'' due to the diverging potential in polar LAO \cite{nakagawa}, is often taken as ``reference''. In fact, the experimentally determined mobile carrier density in LAO/STO is always much smaller than that.
Typical experimental values reported for the 2D-carrier density, $n_s$, measured by Hall effect at 100 K are between 10$^{13}$ and 10$^{14}$ e/cm$^{2}$ \cite{gariglio}. This suggests an important partial charge localization or other charge-compensating mechanisms such as surface passivation or reconstruction \cite{Xie2011, Bristowe}. 

Direct access to the electronic band structure is a crucial step towards the full understanding of complex-oxide interfaces. Angle-resolved photoemission spectroscopy (ARPES) is a powerful technique which yields a map of photoelectron intensities as a function of their kinetic energies and momentum, revealing the electronic structure in solids. Recent ARPES studies on bare STO surfaces \cite{meevasana, santander, PlumbSTO} have shown the formation of a 2DES which appears to have similar properties to those of the LAO/STO interfacial conducting layer. Photoemission spectroscopy of buried interfaces is more challenging due to the small inelastic electron mean free path in solids, which, over a wide range of photon energies ($h\nu$), is of the order of 1~nm. However, it was shown recently that a combination of soft x-ray photoemission with resonant photoexcitation \cite{Kobayashi,Drera, Koitzsch, Cancellieri2013, Claessen} can overcome this limitation. By selecting $h\nu$ at the Ti $L$ edge, the signal of the Fermi states  associated with conducting electrons is greatly enhanced in LAO/STO interfaces. Berner and coworkers \cite{Claessen} were the first to report on the Fermi surface (FS) of the LAO/STO interface, confirming the strong similarities between bare STO and STO-based heterostructures. Here we report the first photoemission measurements resolved in both angle and photon polarization which are thus capable to distinguish different orbital contributions to the bands and FSs. 

We have correlated the FS shape with the number of carriers measured by magnetotransport, investigating samples with different $n_s$. The FS and band dispersions visible in the photoemission experiment are highly dependent on the incident photon light polarizations and on $n_s$, suggesting a strong orbital character and different band filling of these heterostructures. The experimental data are complemented by \textit{ab-initio} results which for the first time describe the detailed evolution of band energies and FS with the charge density present at the interface, in a density range ($\sim 10^{13}$ e/cm$^{2}$) consistent with that measured for the investigated samples.  This provides an unprecedented, direct comparison of calculated and measured electronic properties at equal doping.
Our photoemission band dispersions and theoretical calculations coherently reproduce the orbital decomposition of charges at the given transport-derived carrier density, thus clarifying the exact $3d$ level electron occupancy in this system. The possible contribution of photocarriers induced by x-ray radiation was investigated and only an insignificant effect on the electronic structure was measured. Details are given in the Supplementary Materials.

LAO thin films were grown by pulsed laser deposition on (001)-oriented TiO$_2$-terminated STO substrates at $800\,^{\circ}$C in an oxygen pressure of $8 \times 10^{-5}$~mbar. These ``standard'' LAO/STO samples have transport properties similar to those reported in Ref.~\onlinecite{gariglio}, with $n_s$  $\sim 4-6 \times 10^{13}$ e/cm$^2$, measured by the Hall effect (SD samples). Samples with lower 2D-carriers densities (LD samples) have been also prepared using a growth temperature of 650 $^\circ$C, as reported in Ref.~\onlinecite{caviglia}. These latter samples have a $n_s\sim 10^{13}$ e/cm$^2$. A KrF excimer laser ($248$~nm) was used to ablate the sintered targets with a fluence of $0.6$~J/cm$^2$ at a frequency of $1$~Hz, leading to a deposition rate of about one unit cell for $\sim 60$~pulses. After deposition, the oxygen pressure was raised to $0.2$~bar and the temperature maintained at $540 \pm 10$\,$^{\circ}$C for one hour, in order to ensure full oxidation \cite{Cancellieri}. Film growth was monitored {\it in-situ} by reflection high-energy electron-diffraction. The critical thickness required to undergo an insulator-to-metal transition was $4$~u.c., as reported in Ref.~\onlinecite{thiel06}. Our conducting interfaces had a LAO thickness between $4$ and $5$ u.c.

For \textit{ab-initio} calculations we used an advanced variant of density functional theory, i.e., the variational pseudo self-interaction correction (VPSIC) \cite{vpsic} capable of correcting the imperfect description of standard local-density functionals and reproduce the band-gap of oxide-based systems and the band alignment at the LAO/STO interface \cite{Delugas}. The different carrier density regimes are described starting from the insulating interface (i.e. for LAO thickness smaller than 4 u.c.) and, while keeping the LAO thickness fixed, introducing an increased amount of electron charge in the system, and leaving the system to fully relax to its structural and electronic ground state (in practice mimicking a field-effect-induced charge accumulation). In this way we can monitor the evolution of the electronic properties due to very tiny and well defined changes of electron charge, comparable in magnitude with the Hall-measured values. 

\begin{figure}
\includegraphics[width=10cm]{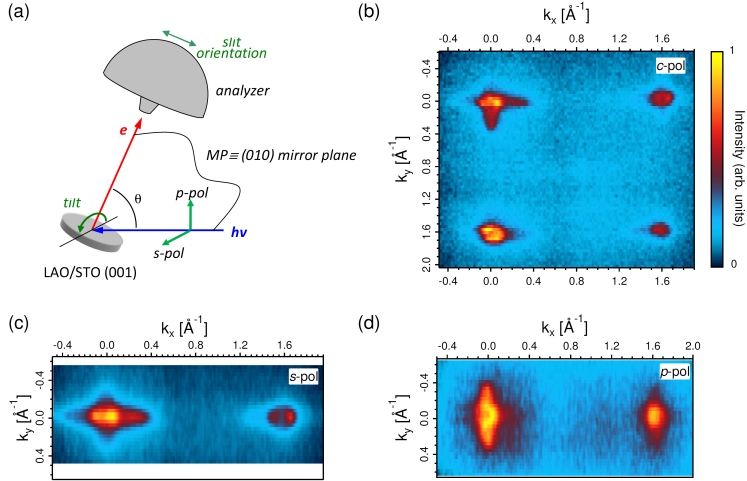}
\caption{(Color online)(a) Experimental geometry setup for polarization dependent measurements; the tilt and $\theta$ angles allow the in-plane $k$-space mapping. (b) FS maps collected for 4 Brillouin zones with circular polarized light at $h\nu$= 460.3~eV; FS maps using (c) $s$-polarized  and (d) $p$-polarized radiation. The data shown are for a SD sample.}
\label{FS}
\end{figure}

The measurements were performed at the soft-x-ray ARPES endstation of the Advanced Resonant Spectroscopies (ADRESS) beamline \cite{ADRESS} at the Swiss Light Source. The samples were transferred from the deposition chamber \textit{ex-situ} without further annealing in vacuum. The beamline delivers a high photon flux, exceeding $10^{13}$~photons/sec/0.01\%~bandwidth providing excellent statistics of the interface signal, despite its attenuation as the photoelectrons pass through the film layer. Measurements were performed using different light polarizations, switching between $c$-polarized (circular), $p$-polarized (linear vertical), and $s$-polarized (linear horizontal), to excite states of different symmetries relative to the mirror plane. The experimental geometry is shown in Fig.~\ref{FS}(a). The measurement plane (MP) coincides with the (010) mirror plane formed by the sample normal along the [001] and the $\Gamma$-X directions. 
The experiment was performed at $\sim11$~K. The combined beamline and analyzer energy resolution was set to around $80$~meV for the valence-band spectra.

Following Ref.~\onlinecite{Cancellieri2013}, $h\nu$ was selected at the Ti $L$ absorption edge (see Supplementary Materials). We maximized the intensity at the Fermi level ($E_{\rm F}$) at sufficient energy separation from the Ti $2p_{3/2}$ core level second order contribution and selected $h\nu$ equal to 460.3~eV. 
We emphasize that no presence of in-gap states could be found in the samples at any $h\nu$ (see Supplementary Materials also for results on the related La$_{0.5}$Sr$_{0.5}$Al$_{0.5}$Ti$_{0.5}$O$_3$/STO heterostructure \cite{Reinle-Schmitt}). These in-gap states, measured in bare STO \cite{meevasana, PlumbSTO} and very recently also in the LAO/STO interface \cite{Claessen, Milan}, appear at $\sim$1~eV below the Fermi edge, are resonant with Ti$^{3+}$ valence states, and are mainly due to surface-defects and impurity states related to oxygen vacancies \cite{Aiura}. These incoherent states are highly photosensitive, displaying an important photodoping effect. Our oxygen-annealed samples show no such contributions from the oxygen vacancies to the 2D electrical conduction, confirmed by the absence of photodoping.
The ARPES data obtained with $c$-polarized photons for a standard LAO/STO sample shown in Fig.~\ref{FS}(b) reveal non-equivalent shapes of the FS in different Brillouin zones, in agreement with Ref.~\onlinecite{Claessen}. This behavior can be ascribed to different matrix elements acting on the photoemission intensity from different interface bands, revealing the compound character of the LAO/STO interface states. The linear dichroism of the spectra, shown in Figs.~\ref{FS}(c) and (d), show more clearly and in detail the multiple-state character of these electronic FS components, which are strongly polarization-dependent. The different photon-polarization measurements reveal that the FS is made up of a circle, originating mainly from Ti $3d_{xy}$ bands, and two ellipsoids aligned along the $k_x$ and $k_y$ directions, due to the heavy $d_{yz}$ and $d_{xz}$ band contributions. A comparison between the $p$- and $s$-polarized data in Fig.~\ref{FS} agrees with the $3d$ orbital character of the FS. The intensity of each band is modulated by the matrix element (photon excitation probability) which depends on the symmetry of the band, the polarization of the incident photons, and their relative angle. 

By switching the incident light polarization from $p$- to $s$-polarization, states symmetric and antisymmetric, respectively, with respect to a (010) MP mirror reflection are excited, thereby selecting the different symmetries of the valence states.
Thus, with $s$-polarization, only the  $d_{xy}$ and $d_{yz}$ states of odd symmetry relative to the MP are detected, while, for $p$-polarized light, only $d_{xz}$ bands with even symmetry with respect to the mirror plane are observed. In Fig.~\ref{FS}(c) ($s$-polarization) the cigar-shaped $d_{yz}$ FS is clearly visible, stretched along the $k_x$ axis, and superimposed on the circular-shaped $d_{xy}$ FS. In Fig.~\ref{FS}(d) ($p$-polarized) only $d_{xz}$ is visible, recognizable by the long side stretched along $k_y$. 

\begin{figure}[!ht]
\centering
\includegraphics[width=10cm]{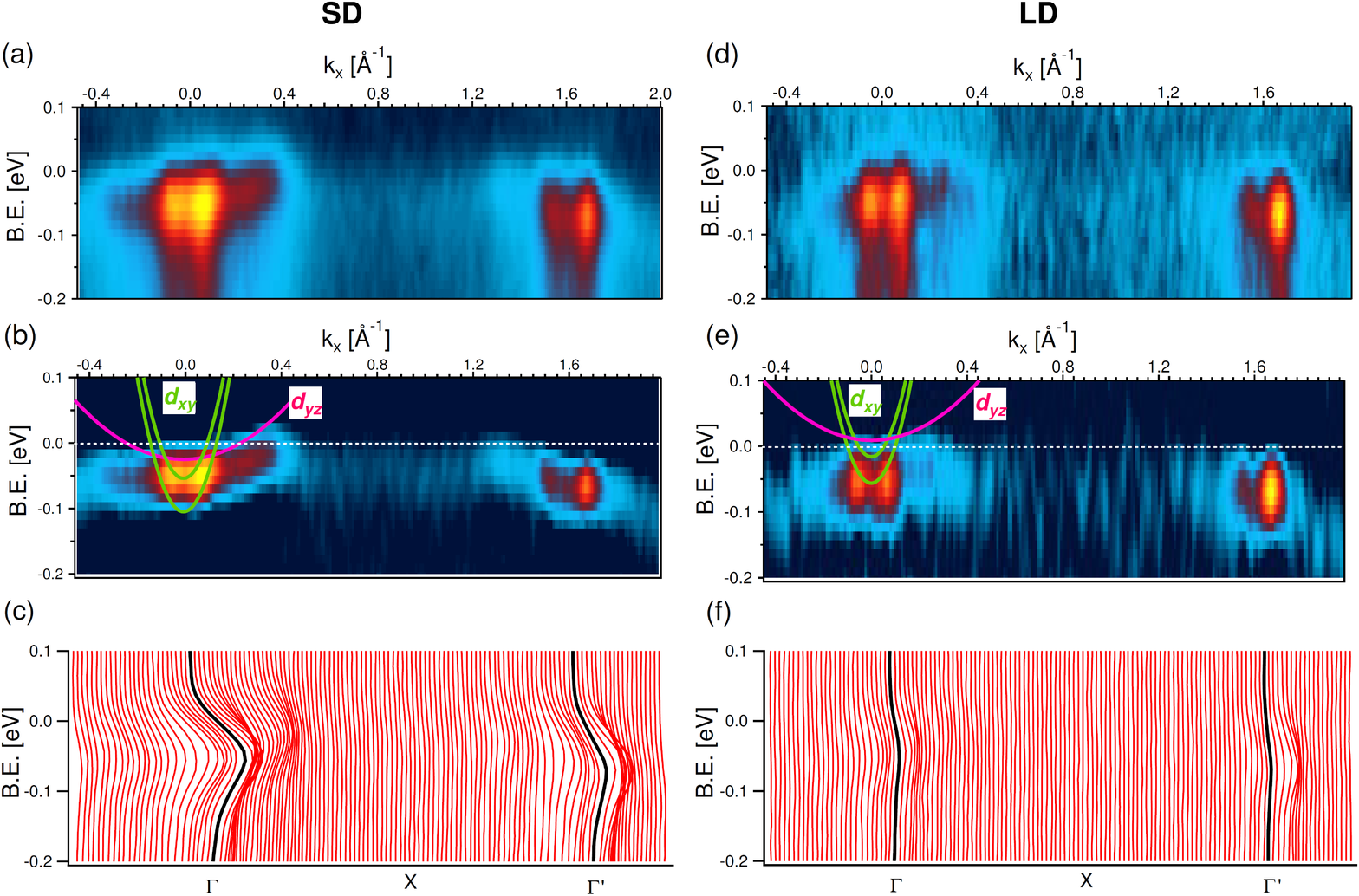}
\caption{(Color online) (a) ARPES data along the $\Gamma$-X-$\Gamma'$ direction with $s$-polarized light and $h\nu= 460.3$~eV collected for a SD sample; (b) Second derivative intensity plot of data in panel (a) illustrates more clearly the band dispersion; (c) EDCs of the data in panel (a). (d)-(f) The equivalent measurements as shown in panels (a), (b) and (c) respectively, but for the LD sample. In panels (b) and (e) the calculated $t_{2g}$ parabolic bands are shown. The color scale is the same as that shown in Fig.~1(b).}
\label{SecondDerHP}
\end{figure}

\begin{figure}[!ht]
\centering
\includegraphics[width=10cm]{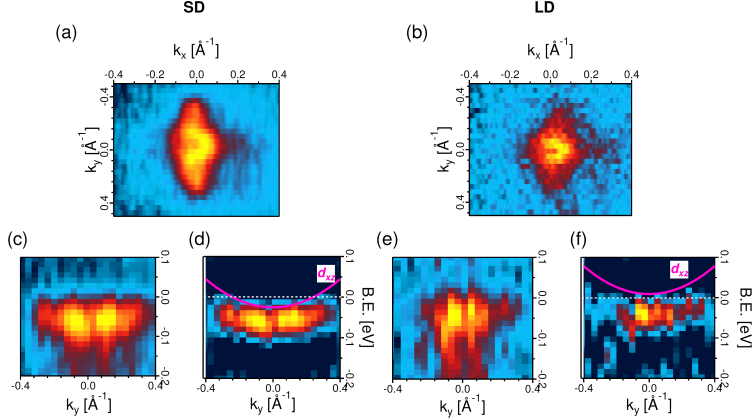}
\caption{(Color online) FS collected with $p$-polarized light and $h\nu=460.3$~eV for (a) a SD and for (b) a LD sample. (c) ARPES data along the $\Gamma$-X direction with $p$-polarized light collected for a SD sample; (d) second derivative intensity plot of data in panel (c) illustrates more clearly the band dispersion; (e) and (f) the equivalent measurements as shown in panels (c) and(d) respectively, but for the LD sample. In panels (d) and (f) the calculated $t_{2g}$ bands are superimposed. The color scale is the same as that given in Fig.~1(b).}
\label{SecondDerVP}
\end{figure}

Fig.~\ref{SecondDerHP}(a)-(f) and Fig.~\ref{SecondDerVP}(a)-(f) show $s$-polarized and $p$-polarized photoemission data respectively, collected around the $\Gamma$ point along the $\Gamma$-X-$\Gamma'$ direction, for both SD and LD samples. Measurements on a LD sample results in a different band dispersion and FS as shown in Fig.~\ref{SecondDerVP}(b): the $d_{xz}$ orbitals are almost no longer visible, highlighting the effect of the doping level on the filling of the bands. The different dispersion signals in the intensity plot Figs.~\ref{SecondDerHP}(a), \ref{SecondDerVP}(c) and (e), and shown more clearly in the negative second derivative plots of Figs.~\ref{SecondDerHP}(b), \ref{SecondDerVP}(d) and (f), demonstrate that different bands are enhanced or suppressed by the polarization geometry.  

In Fig.~\ref{SecondDerHP} are presented the photoemission data acquired with $s$-polarized light. The experimental broadening of the signal at this $h\nu$ precludes the possibility of resolving each individual Ti $3d$ band contributing to the measured photoemission. However, from the photoemission intensity maps, the second derivative plots and the energy-distribution curves (EDCs) in Fig.~\ref{SecondDerHP}, two types of bands, one lighter and one heavier (the latter having lower intensity, thus not visible in the second Brillouin zones), can be clearly distinguished. Comparison with the VPSIC-calculated symmetries of the $t_{2g}$ bands, superimposed on the photoemission data in Fig.~\ref{SecondDerHP}(b) and (e), aids recognition of the orbital character of these bands. For the SD sample [Fig.~\ref{SecondDerHP}(a)] we consider calculation for $n_s$=6.5$\times$10$^{13}$~e/cm$^{2}$, at this density most of the charge is included in three bands. The lowest two bands have planar $d_{xy}$ orbital characters, effective masses m$^*$ $\sim$ 0.7 along $k_x$ and Fermi vectors $k_{\rm F}=0.13$ \AA$^{-1}$ and 0.09 \AA$^{-1}$. They collect the electron charge entirely confined within the first and the second TiO$_2$ layer from the interface (see the Supplementary Material for the detailed analysis). The third occupied band has $d_{yz}$ orbital character and includes charges spreading orthogonally to the interface. It is rather flat along $k_x$ (m$^*$ $\sim$ 9 and $k_{\rm F}= 0.23$ \AA$^{-1}$ according to the calculations) and shifted upward in energy by about 70~meV with respect to the most occupied $d_{xy}$ state. 

In Fig.~\ref{SecondDerHP}(e) (LD sample), we superimposed the photoemission intensity maps with bands calculated for $n_s$=2.6$\times$10$^{13}$~e/cm$^{2}$. Now, $d_{yz}$ is above $E_{\rm F}$ and empty, thus all the charge is included in the two $d_{xy}$ bands, enclosed in circular Fermi pockets of radius $k_{\rm F}$=0.1 \AA$^{-1}$ and 0.06 \AA$^{-1}$. In agreement with the band calculations, the photoemission signal is incompatible with the $d_{yz}$ band.

The $p$-polarization results are shown in Fig.~\ref{SecondDerVP}. By symmetry, only the $d_{xz}$ band are expected to appear, thus here we superimpose only the calculated $d_{xz}$ band [magenta curve in Fig.~\ref{SecondDerVP}(d) and (f)]. Due to tetragonal symmetry, the $d_{xz}$ band bottom is degenerate with $d_{yz}$, and its effective mass along $k_y$ is equal to that of $d_{yz}$ along $k_x$. Thus, for the SD sample, photoemission displays a rather flat band enclosed in a cigar-shaped Fermi pocket of length $k_y\sim 0.3$ \AA$^{-1}$, in good agreement with the value 0.23 \AA$^{-1}$ calculated for $n_s$=6.5$\times$10$^{13}$~e/cm$^{2}$. On the other hand, assuming again $n_s$=2.6$\times$10$^{13}$~e/cm$^{2}$ for the analysis of the LD sample, the calculated $d_{xz}$ band is completely above $E_{\rm F}$. Some $d_{xy}$ admixture shows up in the experiment because of finite acceptance of the analyzer, in the vicinity of the mirror plane.

The interpretation of these results can be understood from the band energies' theoretical evolution as a function of carrier density. First, it is important to realize that, while the in-plane effective masses of the $t_{2g}$ bands are scarcely affected by the interface, their band bottom alignment with respect to $E_{\rm F}$, as well as their mutual band splitting, are crucially dependent on the overall amount of mobile charge in the system, as clearly shown in Fig.~\ref{split} (a detailed description is presented in the Supplementary Materials). Thus, two samples with substantially different doping may present different charge redistribution in individual orbitals, and in turn radically different photoemission response. At very low density, mobile charge accumulates in the $d_{xy}$ bands closer to the interface [the two $d_{xy}$ bottom energies in Fig.~\ref{split} clearly correspond to the green bands shown in Figs.~\ref{SecondDerHP}(b) and (e)], while orthogonal $d_{yz}$ and $d_{xz}$ orbitals [magenta lines in Figs.~\ref{SecondDerHP}(b), (e) and \ref{SecondDerVP} (d), (f)] are initially empty. Such a $d_{xy}$-($d_{yz}$,$d_{xz}$) on-site splitting is a well known consequence of the conduction band misalignment and the suppression of the interface orthogonal hopping. For increasing charge, all bands progressively shift downward in energy across $E_{\rm F}$, until, above a certain charge threshold $n_s$ $\sim$ 3.5 $\times$10$^{13}$~e/cm$^{2}$ (see Fig.~\ref{split}), a portion of the charge starts to fill the orthogonal (cigar-shaped) orbitals as well. Given the different character of planar and orthogonal orbitals, crossing this threshold represents a dramatic change in optical and transport properties of the 2DES. Our photoemission data for two samples purposely prepared for having doping concentration lower and higher than the theoretically predicted threshold, demonstrates explicitly this regime change. The band bottom energies continue to evolve strongly at increasing carrier concentrations: at $n_s =3.3 \times 10^{14}$ e/cm$^2$ corresponding to the half-electron per interface u.c., the $d_{xy}$ (Ti$_1$), $d_{xy}$ (Ti$_2$) and $d_{xz}$ band bottoms are located at -370, -210 and -100~meV, respectively \cite{Delugas}. These values are clearly incompatible with the experimental observations, demonstrating that the ideal limit dictated by the polarization catastrophe is not achieved in actual samples. 
These results differ from the ARPES measurements on bare STO - in contrast 
to the universal 2D FS of bare STO, whose size and shape are decoupled from
the bulk sample preparation, it appears that the deposition of LAO with
different growth conditions affects the LAO/STO 2D FS size. Moreover our dispersion curves and FSs show that the carriers involved in conductivity have a coherent character and the incoherent part (i.e. in-gap states) is negligible.

\begin{figure}
\centering
\includegraphics[width=10cm]{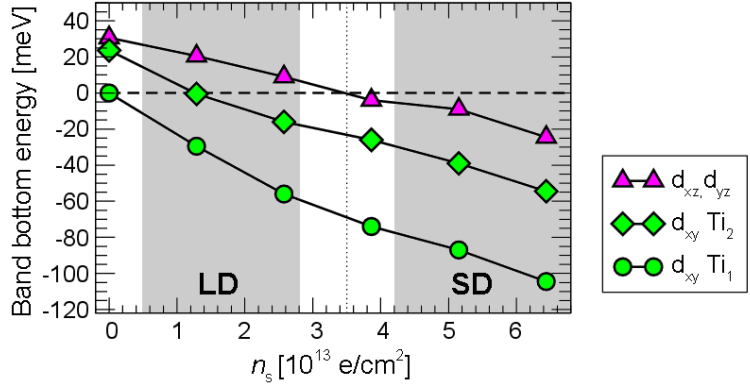}
\caption{(Color online) Band bottom energies with respect to $E_{\rm F}$=0, as a function of the 2D carrier concentration $n_s$ for the most important $t_{2g}$ conduction bands, calculated by VPSIC. First and second lowest interface-parallel $d_{xy}$ bands reside in the first and second TiO$_2$ layer from the interface, respectively. The dotted vertical line indicates the threshold density above which the interface-orthogonal ($d_{xz}$, $d_{yz}$) states  start to contribute to photoemission and transport. Blue shaded areas span the Hall-estimated range of charge density values for LD and SD samples.} 
\label{split}
\end{figure}

In conclusion, we have carried out a polarization-dependent ARPES
experiment investigating the symmetry of the valence states and band
structure of LAO/STO interfaces. We identify
clearly $d_{xy}$, $d_{yz}$ and $d_{xz}$ bands and directly compare these 
with \textit{ab-initio} calculations performed for different band
symmetries. Two types of samples have been considered, one having
a standard, the other a lower charge density. From first-principles calculations, we
have shown that the position of the $3d$ levels is strongly dependent on
the band occupancy.
For both samples, ARPES spectra are compatible with the electronic band
structures calculated for doping concentrations fixed to the values 
obtained from Hall measurements, but not with the band structure
calculated at the ideal limit of $0.5$~e/u.c. expected in the polar catastrophe scenario. This demonstrates that Hall
measurements properly probe the occupancy of the delocalized $3d$ levels
at the interface. This work thus provides benchmark results linking the
evolution of the band structure to the sheet carrier density.
\begin{acknowledgments}
The authors are grateful to J.-M. Triscone, S. Gariglio, D. Stornaiuolo and A. F\^ete for discussion and help in sample preparation.
Support of this work by the Schweizerischer Nationalfonds zur F\"orderung der wissenschaftlichen Forschung, in particular the National Center of Competence in Research, Materials with Novel Electronic Properties, MaNEP. The staff of the Swiss Light Source is gratefully acknowledged. CNR-IOM scientists acknowledge MIUR-PRIN 2010 {\it Oxide}, IIT-Seed NEWDFESCM, IIT-SEED POLYPHEMO and ``platform computation" of IIT, and Fondazione Banco di Sardegna grants. PhG acknowledges a Research Professorship of the Francqui Foundation and partial financial support from the ARC project TheMoTherm.

\end{acknowledgments}
\bibliography{PRL_2012-06}

\end{document}